\documentclass[jkps,preprint,fleqn,showpacs,showkeys]{revtex4}
\usepackage[toc,page]{appendix}
\usepackage[pdftex]{graphicx}
\usepackage{amssymb}
\usepackage{amsmath}
\usepackage{bm}

\begin{document}
\setcounter{page}{0}
\title[]{Development of a free boundary Tokamak Equilibrium Solver (TES) for Advanced Study of Tokamak Equilibria}
\author{YoungMu \surname{Jeon}}
\email{ymjeon@nfri.re.kr}
\thanks{Fax: +82-42-879-5127}
\affiliation{National Fusion Research Institute, Daejeon, Korea
305-333}

\date{\today}

\begin{abstract}
A free-boundary Tokamak Equilibrium Solver (TES), developed for advanced study of tokamak equilibra, is described with two distinctive features.
One is a generalized method to resolve the intrinsic axisymmetric instability, which is encountered after all in equilibrium calculation with a free-boundary condition.
The other is an extension to deal with a new divertor geometry such as snowflake or X divertors.
For validations, the uniqueness of a solution is confirmed by the independence on variations of computational domain, the mathematical correctness and accuracy of equilibrium profiles are checked by a direct comparison with an analytic equilibrium known as a generalized Solov'ev equilibrium, and the governing force balance relation is tested by examining the intrinsic axisymmetric instabilities. 
As a valuable application, a snowflake equilibrium that requires a second order zero of the poloidal magnetic field is discussed in the circumstance of KSTAR coil system.
\end{abstract}



\maketitle

\section{Introduction}

In tokamak physics, plasma equilibrium is a fundamental and essential element to understand not only the basic equilibrium properties but also various plasma phenomena such as MHD instabilities, plasma transport and turbulence, plasma flows and waves, and so on.
Therefore, various numerical or analytical equilibrium studies \cite{Takeda1991JCP} have been conducted for a long time since the axisymmetric plasma equilibrium relation was established in a general form, known as Grad-Shafranov equation \cite{Shafranov1958, Grad1958}.

Depending on the characteristics of applications, the studies can be categorized into two types of problems.
One, so called 'fixed boundary equilibrium', is solving an equilibrium assuming that the plasma boundary or plasma region is known.
So, the external equilibrium field is ignored and the internal equilibrium profiles and flux distributions are mainly concerned.
The other one, so called 'free boundary equilibrium', is solving the equilibrium with unknown plasma boundary.
Hence the plasma position and shape (i.e. plasma region) need to be obtained as a solution, in addition to those equilibrium profiles and flux distributions.

Due to the importance of equilibrium as a basis for various physics studies, the majority of equilibrium studies has been devoted to the fixed boundary equilibrium problems, while less interests given to the free boundary equilibrium solutions. 
However, recently new demands for the free boundary equilibrium analysis have been arisen and turned out to be important.
For instance, another type of equilibria with new topological features, so called snow-flake (SF) divertor \cite{Ryutov2007PoP} or X-divertor \cite{Kotschen2007PoP} equilibria, have been proposed and actively studied in various devices \cite{Piras2009PPCF,Souk2011NF} recently.
Specially, since the SF divertor configuration requires a second-order zero of the poloidal magnetic field, it is now an important issue that should be addressed in terms of a free boundary tokamak equilibrium \cite{Lackner2013FST}.

Accordingly, a free boundary Tokamak Equilibrium Solver, called as TES, has been developed with an emphasis on applications to a design work of plasma equilibrium control and to advanced equilibrium study.
The developed TES code is featured by two distinctive functionalities; a generalized method for stabilization of axisymmetric instabilities, and an extension to deal with a second-order zero of the poloidal magnetic field.

In section \ref{sec:solution_method}, the numerical solution methods and procedures used in TES is described for two types of free boundary equilibrium problems, i.e., {\it ideally free} and {\it semi free} boundary problems that will be defined therein.
Most of numerical techniques and issues have been well known, so that only a brief description for each issues is given unless necessary.
For validations of TES code, a direct comparison with a generalized analytic solution is described in section \ref{sec:validation}, in addition to the uniqueness of a solution.
In section \ref{sec:stability}, an intrinsic axisymmetric instability, encountered after all during the numerical procedure, is tested by examining the variations of plasma equilibria, and a generalized stabilization method is introduced and tested.
In section \ref{sec:extension}, an extended feature to deal with a snowflake divertor is explained and discussed, followed by a summary and conclusion in section \ref{sec:conclusion}.

\section{A solution method in TES}
\label{sec:solution_method}

A basic numerical method and procedure of an axisymmetric tokamak plasma equilibrium with a free boundary condition has been well established \cite{Johnson1979JCP, Jardin1986JCP, Hofmann1988CPC}.
The mathematical and numerical treatments used in TES code is also basically in line with those in the references, except some improved and extended features.
Therefore, in this section, we describes the basic numerical treatments and procedures used in TES code briefly unless necessary.

\subsection{Force balance relation for free boundary plasma equilibrium}
In a toroidally axisymmetric system like a tokamak, the force balance relation of a plasma, i.e. plasma equilibrium, can be expressed by a second-order partial differential equation, known as Grad-Shafranov equation \cite{Shafranov1958, Grad1958}, using a cylindrical coordinate system $(R,\phi,Z)$ with an ignorable (due to axisymmetry) toroidal angle coordinate $\phi$.
\begin{eqnarray}
\label{eq:Ampere}
\triangle ^*\psi(R,Z) &=&  -\mu_0 R J_{\phi,pl} (R,Z) \\
\label{eq:GS}
J_{\phi,pl}(R,Z) &=& Rp'(\psi) + \dfrac{ F(\psi)F(\psi)' } {\mu_0 R }
\end{eqnarray}
where the poloidal flux function (equal to the actual poloidal magnetic flux divided by $2\pi$) is defined by $\psi(R,Z)\equiv RA_{\phi}(R,Z)$ from $B=\nabla \times A$ (i.e. $\nabla \cdot B=0$), the Shafranov operator defined by $\triangle ^* \equiv R^2 \nabla \cdot \genfrac{}{}{}{0} {\nabla}{R^2}$, and the prime denotes $g' \equiv \genfrac{}{}{}{0} {\partial g}{\partial \psi}$.
And $p(\psi)$ is an isotropic plasma pressure, $F(\psi) \equiv RB_{\phi}$ a toroidal field function, and $J_{\phi,pl}(R,Z)$ a toroidal current density of plasma.
Since the $J_{\phi,pl}(R,Z)$ as a source term in Eq. (\ref{eq:Ampere}) has a strong dependency on $\psi(R,Z)$ by Eq. (\ref{eq:GS}), it gives rise to a strong non-linearity on the equations.

In order to deal with a free boundary condition in equilibrium calculation, the Eq. (\ref{eq:Ampere}) is generalized by including arbitrary toroidal conductor currents as follows.
\begin{eqnarray}
\label{eq:GSe}
\triangle ^*\psi(R,Z) &=&  -\mu_0 R J_{\phi} (R,Z) \\
J_{\phi}(R,Z) &=& J_{\phi,pl}(R,Z) + J_{\phi,cond}(R,Z) \nonumber
\end{eqnarray}
where $J_{\phi,cond}(R,Z)$ is the toroidal current density for a conductor.
The toroidal conductor could be any toroidal current source that can affect the equilibrium force balance, such as poloidal field (PF) coil currents or axisymmetric eddy currents on surrounding conductor structures.
Assuming discrete conductors with uniform current distributions inside, the toroidal conductor current density can be expressed by
\begin{eqnarray}
\label{eq:jcond}
J_{\phi,cond}(R,Z) &=& \sum^{N_{cond}}_{k=1}{ J_{cond,k}(R,Z) } \\
J_{cond,k}(R,Z) &=&
\begin{cases}
I_{cond,k}/S_{k} & \text{ if } (R,Z) \in    \Omega_{cond,k} \nonumber \\
0                & \text{ Otherwise }
\end{cases}
\end{eqnarray}
where $J_{cond,k}$, $I_{cond,k}$, $S_k$, and $\Omega_{cond,k}$ are the toroidal current density, the toroidal current, the cross-sectional area, and the domain region of k-th conductor, respectively.

Meanwhile, the toroidal current density of plasma $J_{\phi,pl}$ in Eq. (\ref{eq:GS}) can be set into a canonical form \cite{Albanese1998NF} as shown below
\begin{eqnarray}
\label{eq:jpl}
J_{\phi,pl}(R,Z) =
\begin{cases}
\lambda \left[ \beta_0 \dfrac{R}{R_{geo}} + (1-\beta_0) \dfrac{R_{geo}}{R} \right] 
\tilde{j}(\psi, \psi_a,\psi_b) & \text{ if } (R,Z) \in    \Omega_{pl} \\
0                & \text{ Otherwise }
\end{cases}
\end{eqnarray}
with $\tilde{j}(\psi,\psi_a,\psi_b) \equiv \left(1-\psi^{\alpha_m}_s \right)^{\alpha_n} $, where $R_{geo}$ is the major radius as a reference length scale, $\psi_a$ the flux per radian at the plasma magnetic axis, $\psi_b$ the flux per radian at the plasma boundary, and $\tilde{j}$ a suitable profile function.
The $\lambda$ and $\beta_0$ are adjustable variables to satisfy equilibrium constraints which will be discussed later, while the $\alpha_m$ and $\alpha_n$ are input variables specified by users.
Note that $J_{\phi,pl}(R,Z)$ is automatically set to zero at the plasma boundary in this form by using a normalized poloidal flux, $\psi_s \equiv (\psi-\psi_a)/(\psi_b-\psi_a)$.

In short, a free boundary plasma equilibrium can be obtained by solving Eq. (\ref{eq:GSe}) with a toroidal current density specified by Eqs. (\ref{eq:jcond}) and (\ref{eq:jpl}), and corresponding equilibrium profiles such as $p(\psi)$ and $F(\psi)$ can be obtained from Eqs. (\ref{eq:GS}) and (\ref{eq:jpl}).

\subsection{Numerical approximation by discretizations}
The governing equation described above, i.e. Eq. (\ref{eq:GSe}), can be thought as a 2D Poisson's equation in toroidal geometry, so that easily solved using various numerical methods if the source term is known.
For numerical treatments, the equation is converted to a linear algebraic equation by using the centered finite difference method (FDM) \cite{NR} on a rectangular computational domain in ($R,Z$) space,
where the grids, ($R_l,Z_j$), are built by
\begin{eqnarray}
\label{eq:gridrz}
R_l = R_{min}+ \Delta R \times (l-1),~~~
\Delta R \equiv (R_{max}-R_{min})/(N_R-1) \nonumber\\
Z_j = Z_{min}+ \Delta Z \times (j-1),~~~
\Delta Z \equiv (Z_{max}-Z_{min})/(N_Z-1) 
\end{eqnarray}
with $l=1,\cdots,N_R$ and $j=1,\cdots,N_Z$. Then the algebraic equation converted by FDM can be expressed as follows.
\begin{eqnarray}
\label{eq:linearGS}
\frac{1}{(\Delta Z)^2} \psi_{j-1,l}
&+& \left( \frac{1}{(\Delta R)^2} + \frac{1}{2R_l(\Delta R)} \right) \psi_{j,l-1} 
- \left\{ 2 \left( \frac{1}{(\Delta R)^2} + \frac{1}{(\Delta Z)^2} \right) \right\} \psi_{j,l} \nonumber\\
&+& \left( \frac{1}{(\Delta R)^2} - \frac{1}{2R_l(\Delta R)} \right) \psi_{j,l+1} 
+ \frac{1}{(\Delta Z)^2} \psi_{j+1,l} = - \mu_0 R_l J_{\phi j,l}
\end{eqnarray}
where $\psi_{j,l}= \psi(R_l,Z_j)$ and $J_{\phi j,l}= J_{\phi}(R_l,Z_j)$ with $l=2,\cdots,N_R-1$ and $j=2,\cdots,N_Z-1$. 
This algebraic equation can be solved by either using a matrix inversion after reforming it in a form of ${\bf Ax=b}$ or using an iterative method such as multi-grid method \cite{NR} or double cyclic reduction \cite{Buneman}, with an appropriate boundary condition. 
In TES code, the successive-over-relaxation (SOR) method \cite{NR} is used as a basic numerical scheme for the simplicity.

\subsection{Iterative solution for non-linearity}
To solve the Eq. (\ref{eq:linearGS}) in the given form, the source term on the right hand side should be known. 
However, the plasma part of the source term has a strong nonlinear dependency on $\psi_{j,l}$ according to Eq. (\ref{eq:GS}) or (\ref{eq:jpl}).
To deal with this non-linearity, an iteractive method, known as Picard iteration \cite{EngMath}, is adopted.
Then, the Eq. (\ref{eq:linearGS}) is expressed as follows
\begin{eqnarray}
\label{eq:picard}
\frac{1}{(\Delta Z)^2} \psi^{(n)}_{j-1,l}
&+& \left( \frac{1}{(\Delta R)^2} + \frac{1}{2R_l(\Delta R)} \right) \psi^{(n)}_{j,l-1} 
- \left\{ 2 \left( \frac{1}{(\Delta R)^2} + \frac{1}{(\Delta Z)^2} \right) \right\} \psi^{(n)}_{j,l} \nonumber\\
&+& \left( \frac{1}{(\Delta R)^2} - \frac{1}{2R_l(\Delta R)} \right) \psi^{(n)}_{j,l+1} 
+ \frac{1}{(\Delta Z)^2} \psi^{(n)}_{j+1,l} \\
&=& - \mu_0 R_l J^{(n)}_{\phi j,l} \left( \psi^{(n-1)}_{j,l} \right) \nonumber
\end{eqnarray}
where $(n)$ indicates the n-th Picard iteration.
Note that the source term in the n-th iteration, $J^{(n)}_{\phi j,l}$, is expressed as a function of $\psi^{(n-1)}_{j,l}$, i.e. the poloidal flux in the (n-1)th iteration.
Hence, the $\psi^{(n)}$ is obtained from Eq. (\ref{eq:picard}) using $J^{(n)}_{\phi}$ that was evaluated from $\psi^{(n-1)}$. 
Then, the $J^{(n+1)}_{\phi}$ is updated using the refreshed $\psi^{(n)}$ and provided as a new source term into Eq. (\ref{eq:picard}).
This recursive iteration is continued until a convergence criterion, $\left\| \psi^{(n)}-\psi^{(n-1)} \right \| < \epsilon$, is satisfied.

\subsection{Boundary conditions}
In general, the boundary condition in free boundary equilibrium calculation is not constant and varied due to changes of plasma boundary and equilibrium profiles during the numerical iterations, while in a fixed boundary equilibrium it is fixed to zero ($\psi_{bc}=0$) usually.
The Dirichlet boundary condition on the edge of a computational domain can be provided directly by using a Green's function formulation \cite{Miyamoto}.
\begin{eqnarray}
\psi(R,Z)= \iint {G(R,Z;R',Z')J_{\phi}(R',Z')}dR'dZ' 
\end{eqnarray}
where $G(R,Z;R',Z')$ is the free space Green's function which gives the poloidal flux at $(R,Z)$ from a unit toroidal current source at $(R',Z')$.
The free space Green's function is defined by 
\begin{eqnarray}
G(R,Z;R',Z')&=&\dfrac{\mu_0}{2\pi}\dfrac{\sqrt{RR'}}{k}
\left[ \left( 2-k^2 \right ) K(k) - 2E(k)
\right] \nonumber\\
k^2 &\equiv & \dfrac{4RR'}{ \left( R+R' \right)^2 + \left( Z-Z' \right)^2 }
\end{eqnarray}
where $K(k)$ and $E(k)$ are elliptic integrals of the first and the second kind \cite{NR}, respectively.
Using this, the poloidal flux at the boundary of computational domain can be directly obtained by taking into account both plasma and conductor currents as follows
\begin{eqnarray}
\psi_{bndry}^{(n)}(R_b,Z_b) &=& \int_{\Omega_{pl}'} {G(R_b,Z_b;R',Z')J^{(n)}_{\phi,pl}(R',Z')}d\Omega_{pl}'  \nonumber\\
&+&\int_{\Omega_{cond}'} {G(R_b,Z_b;R',Z')J^{(n)}_{\phi,cond}(R',Z')}d\Omega_{cond}'
\end{eqnarray}
where $(R_b,Z_b)$ is the boundary point of the computational domain.
Note that $J^{(n)}_{\phi,pl}(R',Z')$ is varied in every steps of Picard iterations, while $J^{(n)}_{\phi,cond}(R',Z')$ is not changed unless the plasma boundary is specified, which will be discussed later.

\subsection{Determination of plasma boundary}

For a stable convergence of the solution, it is important to accurately determine the plasma region or boundary in terms of $\psi_b$ in every steps of Picard iteration.
Generally a plasma boundary is formed either by limiters (a limited plasma) or by magnetic fields with an X-point (a diverted plasma).
Assuming $I_p>0$, the poloidal flux, $\psi(R,Z)$, has a convex distribution inside plasma, thus $\psi_a>\psi_b$. 
Therefore, the $\psi_b$ is defined by the maximum value among all poloidal fluxes from limiters and from X-points.
When $I_p<0$, the $\psi_b$ is defined by the minimum value in a same logic.

More precisely, both magnetic axis and X-point have a null-field ($\lvert \nabla \psi \rvert ^2=0$), while they have different signs of second-derivatives \cite{Johnson1979JCP}, defined by
\begin{equation}
S(R,Z) \equiv
  \left( \genfrac{}{}{}{}{\partial^2 \psi}{\partial R^2} \right)
  \left( \genfrac{}{}{}{}{\partial^2 \psi}{\partial Z^2} \right)
 -\left( \genfrac{}{}{}{}{\partial^2 \psi}{\partial R \partial Z} \right)^2
\end{equation}
If $S>0$, the field-null point is a magnetic axis ($\psi_a$). Otherwise ($S<0$), it is an X-point.
The accurate location of the magnetic axis or the X-point is determined by using the Powell's conjugate direction method \cite{NR} based on a 2D bicubic interpolation.

\subsection{Constraints on plasma equilibrium}
\label{eq_constraint}

In order to have a unique equilibrium solution for Eq. (\ref{eq:GS}), a few constraints on plasma equilibrium quantaties are necessary. 
Considering the functional form of Eq. (\ref{eq:jpl}), two constraints, total plasma currents and poloidal plasma beta, are applied.
Note that the equilibrium constraints could be different when a different functional form of $J_{\phi,pl}(R,Z)$ is used instead of Eq.(\ref{eq:jpl}).
For instance, if $q(\psi)$ profile is used in $J_{\phi,pl}(R,Z)$, then $q_a=q(\psi_a)$ could be used as another appropriate constraint \cite{LoDestro1994PoP}.

The constraints can be expressed as
\begin{subequations}
\begin{align}
\label{eq:ip}
I_p =& \mathop{ \int }_{ \Omega_{pl} } J_{\phi,pl}(\lambda, \beta_0)d\Omega \\
\label{eq:betap}
\beta_p =& \frac{ \langle p(\beta_0) \rangle }{ \langle B_p^2 \rangle _{\psi_a} \slash {2\mu_0} }
\end{align}
\end{subequations}
where $\mu_0=4\pi \times 10^{-7}[N/A^2]$ is the permeability of vacuum, $a$ the minor radius, and $B_p$ the poloidal magnetic field.
The braket $\langle \cdot \rangle$ means an average over a magnetic surface.
Therefore, by combining these two equations, the $\beta_0$ and $\lambda$ can be determined thus giving a unique solution.

\subsection{External equilibrium fields with specified plasma boundary}
\label{sec:fixedEQ}

In principle, for a free boundary equilibrium problem, the plasma boundary is solved as a part of solutions under given external equilibrium fields.
In practice, however, it is more useful and convenient to solve the equilibrium with a specified plasma boundary.
In this study, we distinguish them by calling the former as an {\it ideally free} boundary problem while the latter by a {\it semi free} boundary problem.
In the case of {\it semi-free} boundary, the external coil currents are adjusted to provide a required equilibrium field.
If a plasma boundary is specified in a series of points, the required external equilibrium field currents can be determined by solving a minimization problem as shown below
\begin{eqnarray}
\label{eq:eq_pf}
\min_{\Delta I_{\text{coil}}}
\left[
\sum^{N_{\text{bndry}}}_{j=1} { \left\lbrace \sum^{N_{\text{coil}}}_{i=1} 
  \Big( G(R_j,Z_j;R_i,Z_i)\cdot \Delta I_{\text{coil},i} \Big)
 -\Delta \psi(R_j,Z_j) \right\rbrace^2 } \right. & \nonumber \\
+\sum^{N_{\text{Xpt}}}_{j=1} { \left\lbrace \sum^{N_{\text{coil}}}_{i=1} 
  \Big( G_{B_R}(R_j,Z_j;R_i,Z_i)\cdot \Delta I_{\text{coil},i} \Big)
 -B_{R}(R_j,Z_j) \right\rbrace^2 } & \nonumber \\
+\sum^{N_{\text{Xpt}}}_{j=1} { \left\lbrace \sum^{N_{\text{coil}}}_{i=1} 
  \Big( G_{B_Z}(R_j,Z_j;R_i,Z_i)\cdot \Delta I_{\text{coil},i} \Big)
 -B_{Z}(R_j,Z_j) \right\rbrace^2 } & \nonumber \\
\left.
+\gamma^2 \sum^{N_{\text{coil}}}_{i=1}{ \Big( \Delta I_{\text{coil},i} \Big)^2 }
\right]
\end{eqnarray}
where ($R_j,Z_j$) is the specified j-th boundary point, $\Delta \psi(R_j,Z_j)=\psi_b - \psi(R_j,Z_j)$ is the poloidal flux error on the point, $B_{R}(R_j,Z_j)$ and $B_{Z}(R_j,Z_j)$ are the radial and vertical magnetic fields there, and $\gamma$ is a Tikhonov parameter for regularization \cite{Tikhonov1977}.
If an X-point is specified as a part of plasma boundary, then the radial and vertical magnetic fields at the point should be zeros.
This constraint is added as the second and third terms in Eq.(\ref{eq:eq_pf}) with $G_{B_R} \equiv -\dfrac{1}{R}\dfrac{\partial G}{\partial Z}$ and $G_{B_Z} \equiv +\dfrac{1}{R}\dfrac{\partial G}{\partial R}$.
From this, the external equilibrium field currents are obtained by $I^{(n)}_{\text{coil},j}=I^{(n-1)}_{\text{coil},j}+ \Delta I_{\text{coil},j}$, where $I^{(n-1)}_{\text{coil},j}$ is the coil currents in ($n-1$)th Picard iteration.

\section{Validations of TES}
\label{sec:validation}

According to the numerical methods and procedures described above, a free boundary tokamak equilibrium solver (TES) has been developed. 
For the validations of this code, the uniqueness of a solution is firstly checked by examining the independence on the variations of computation domains, and the mathematical correctness and accuracy of equilibrium profiles are assessed by a direct comparison with an analytic equilibrium solution.

\begin{figure}[!hbtp]
\centering
\includegraphics[width=0.30\textwidth]{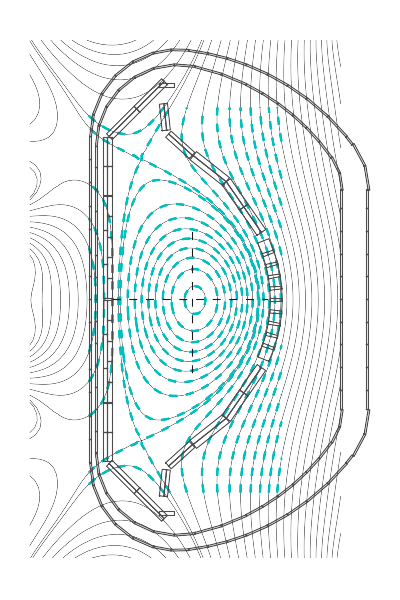}
\includegraphics[width=0.53\textwidth]{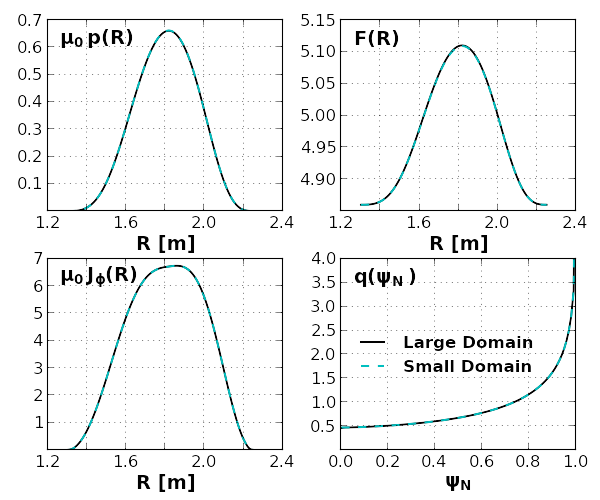}
\caption{(Color online) Two free boundary equilibria, obtained by TES with identical equilibrium constraints, are directly compared.
On the left, the poloidal magnetic flux in a large computational domain (black solid line) is compared with that in a small computational domain (cyon dotted line).
On the right, several equilibrium profiles are directly compared, such as the isotropic pressure, the toroidal field function, the toroidal current density, and the safety factor profile.
}\label{fig:domain}
\end{figure}

\subsection{Uniqueness of a solution under numerical variations}
Since we are solving the problem in a numerical approach, one fundamental test, which is seldom seen in the related literatures, is to examine if it provides an identical result, independent on the number of grids or the change  of computational domain.
Particularly it is essential and critical when a free boundary condition is imposed.

A comparison of two free boundary equilibrium solutions, one in a large and the other in a small computational domains, is shown in Fig. \ref{fig:domain} where $I_p=-2.0$ MA, $B_T=-2.7$ T, $a=0.48$ m, and $\beta_p=0.5$ with a large elongation $\kappa=2.0$.
The poloidal magnetic fluxes are compared on the left and several equilibrium profiles on the right.
The solution for a large computational domain (black solid line) was obtained in $0.7 \le R \le 2.8$ m, $-1.9 \le Z \le +1.9$ m with $N_R \times N_Z=65\times 85$, while the one for a small computational domain (cyon dotted line) in $1.1 \le R \le 2.4$ m, $-1.3 \le Z \le +1.3$ m with $N_R \times N_Z=45\times 65$.
As expected, the poloidal magnetic fluxes and the equilibrium profiles are shown to be almost identical for both.
Therefore it confirms that the equilibrium obtained by TES provides a unique solution, independently on any change of computational domain and the grid size.

\begin{figure}[!hbtp]
\includegraphics[width=5.0cm]{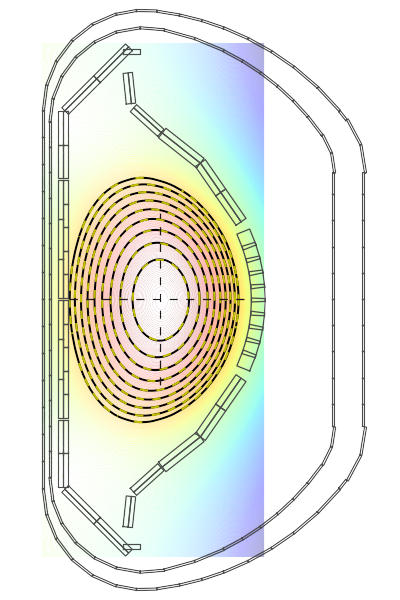}
\caption{(Color online) The poloidal magnetic flux obtained from TES (black solid line) is compared directly with the analytic one from the Solovev's equilibirum (yellow dotted line). The filled contour plot with a rectangular boundary shows the full distribution of poloidal magnetic flux including vacuum region, which is obtained from TES.
}\label{fig:solovev}
\end{figure}

\subsection{Benchmark with an analytic solution}
For a direct validation of TES, an analytic fixed boundary equilibrium solution, known as a generalized Solov'ev equilibrium \cite{Zheng1996PoP}, is considered and compared with a TES result.
Note that it is to check the mathematical correctness and accuracy of the solution from TES.
The pressure and toroidal field function in the analytic solution are assumed to be constant
\begin{equation}
-\mu_0 \frac{ \partial p }{ \partial \psi }=A_1,~~~
F \frac{\partial F}{\partial \psi}=A_2
\label{eq:simple_jphi}
\end{equation}
Then, the equilibrium solution can be expressed explicitly as follows
\begin{equation}
\psi(R,Z)=c_1 + c_2 R^2 + c_3 (R^4-4 R^2 Z^2) + c_4 \left[ R^2 \ln(R)-Z^2 \right] + \frac{R^4}{8} A_1 - \frac{Z^2}{2} A_2
\label{eq:solution}
\end{equation}
where four constants, $c_i,~i=1,\cdots,4$ are determined to satisfy the boundary conditions from specified plasma boundary, and other two parameters, $A_1$ and $A_2$, are adjusted to meet the equilibrium constraints.
Four boundary conditions, with a modification for the comparison, are given by  $\psi(R_{in},Z_{in})=\psi(R_{out},Z_{out})=\psi(R_{top},Z_{top})=\psi_b$ and   $\genfrac{.}{|}{}{} {\partial \psi}{\partial R} _{\left(R_{top},Z_{top}\right)}=0$, where $in$, $out$, and $top$ are the inner-, the outer-, and the top-most boundary points respectively.
The equilibrium constraints are the total plasma currents $I_p$ and the poloidal beta $\beta_p$ as follows
\begin{subequations}
\begin{align}
I_p =& \mathop{ \int }_{ \Omega_{pl} } J_{\phi}dRdZ 
= - \left( \mathop{ \int }_{ \Omega_{pl} } \left( \frac{R}{\mu_0} \right) dRdZ \right) A_1
+ \left( \mathop{ \int }_{ \Omega_{pl} } \left( \frac{1}{\mu_0 R} \right) dRdZ \right) A_2 \\
\beta_p =& \frac{8 \pi}{\mu_0} \frac{ \int p~dRdZ }{I_p^2} 
= \left( \frac{8 \pi}{ (\mu_0 I_p)^2 } \int (\psi_b - \psi)dRdZ \right) A_1 
\label{eq:constraints2}
\end{align}
\end{subequations}
thus determining appropriate values of $A_1$ and $A_2$.

The poloidal magnetic fluxes obtained from TES and the analytic solution are directly compared in Fig. \ref{fig:solovev}, where $I_p=0.5$ MA, $B_T=2.7$ T, $\beta_p=0.5$ with elongation $\kappa=1.45$ and minor radius $a=0.5$ m in a limited configuration (i.e. without null point).
The full distribution (including vacuum region) of poloidal magnetic flux from TES is shown as a filled contour plot with a rectangular boundary, to show it is indeed a free boundary solution.
The poloidal magnetic fluxes in plasma region are directly compared by overlapping them; one is from TES (black solid line) and the other from the analytic solution (yellow dotted line).
As shown, two results are not distinguishable and thus the difference is negligible.
It confirms that the equilibrium informations inside plasma region, obtained by TES, are accurately consistent with those from analytic calculations.

Summarizing two validation results above, it is confirmed that TES provides a unique equilibrium solution with high accuracy, consistent with theoretical analysis.

\section{Axisymmetric instability and its stabilization}
\label{sec:stability}

\subsection{Axisymmetric instability of shaped plasma equilibrium}
A tokamak plasma equilibrium has 2D axisymmetric, instrinsic instabilities associated with plasma shaping.
The most important 2D axisymmetric, i.e. the toroidal mode number n=0, instability is known as a vertical instability which becomes unstable once a plasma elongation is increased above a threshold.
It has been well understood that this instability is originated from a $J_{\phi,pl} \times B_{ext,pol}$ force on plasma by an external equilibrium field due to a bad curvature associated with the plasma shape.
The field curvature can be evaluated by a field decay index, $n_{\text{decay}}$, defined as
\begin{equation}
n_{\text{decay}}(R,Z) \equiv -\dfrac{R}{B_Z}
  \genfrac{}{}{}{}{\partial B_Z}{\partial R}
= -\dfrac{R}{B_Z}
  \genfrac{}{}{}{}{\partial B_R}{\partial Z}
\label{eq:ndecay}
\end{equation}
Theoretically, it is well known that a vertically elongated plasma can be unstable when $n_{\text{decay}} < 0$ and a radially elongated plasma unstable when $n_{\text{decay}} > 3/2$ \cite{Fukuyama1975JJAP}.

\begin{figure}[!hbtp]
\centering
\includegraphics[width=0.22\textwidth]{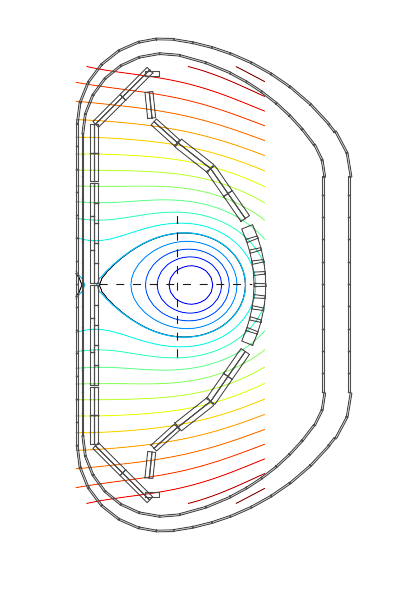}
\includegraphics[width=0.50\textwidth]{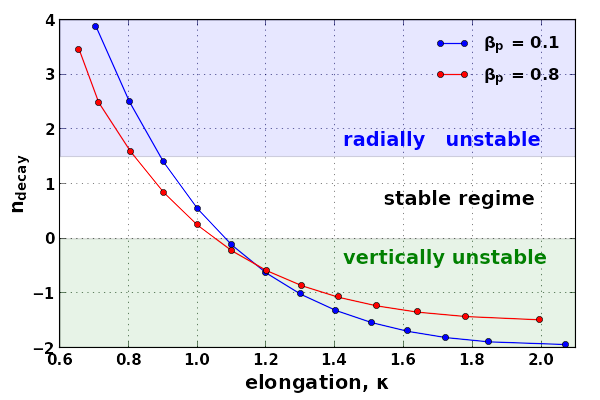}
\includegraphics[width=0.22\textwidth]{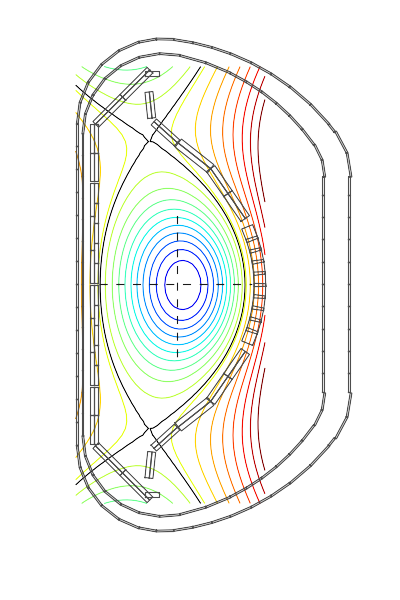}
\caption{(Color online) Decay index of external equilibrium fields ($n_{\text{decay}}$) vs plasma elongation ($\kappa$) for two different plasma betas ($\beta_p$).
Theoretically stable and unstable regimes in terms of $n_{\text{decay}}$ are marked with filled colors.
Additionally a radially elongated plasma with $\kappa=0.7$ (left) and a vertically elongated plasma with $\kappa=2.0$ (right) are shown for $\beta_p=0.8$.
}\label{fig:ndecay}
\end{figure}

The relation between plasma shape and field decay index can be seen in FIG. \ref{fig:ndecay} for a plasma of $I_p=0.5$ MA and $B_T=2.5$ T, where theoretically stable and unstable regimes are marked with filled colors.
Comparison of two plasmas with different beta ($\beta_p$) shows that the plasma with higher $\beta_p$ is less unstable and has wider range of stable $\kappa$, consistently with theory.
Note that the series of equilibria in this figure is obtained by specifying the plasma boundary (i.e. as a {\it semi free} boundary problem), in order to avoid the axisymmetric instability due to the bad curvature.

\subsection{A generalized stabilization for axisymmetric instabilities}
Due to the axisymmetric instability, the direct solution of plasma equilibrium under given external equilibrium field (i.e. as an {\it ideally free} boundary problem) has a convergence issue.
That is, a small deviation of plasma from an equilibrium position is inevitable during a numerical iteration, so that the plasma could be drifted and eventually diverged either radially (when $n_{decay}>1.5$) or vertically (when $n_{decay}<0.0$).
In the literature, a conventional method to resolve the vertical instability of elongated plasma is simply inserting a feedback loop \cite{Johnson1979JCP} by adding artificial feedback coils which are typically a pair of up-down symmetric coils to produce a horizontal magnetic field.
In this method, the feedback coil currents are adjusted to control the vertical position of magnetic axis to a pre-selected target position according to the relation below
\begin{equation}
I_{\text{feedback}}=-\text{sign}(Z_{\text{coil}}) \left\lbrace C_1 (Z^{(n)}_{\text{mag}} - Z_{\text{target}}) + C_2(Z^{(n)}_{\text{mag}}-Z^{(n-1)}_{\text{mag}}) \right\rbrace I_p
\label{eq:fbzp}
\end{equation}
where $Z_{\text{target}}$ is the desired vertical position of the magnetic axis and $Z_{\text{coil}}$ is the vertical position of the control coil. 
A critical drawback of this method is that the desired vertical position of the magnetic axis should be known, prior to obtaining it as a solution from the equilibrium calculation.
Also the constants $C_1$ and $C_2$ are chosen by trial and error.

In TES code, the Eq. (\ref{eq:fbzp}) is modified for a general treatment.
Instead of controlling the vertical position in a feedback manner, we are eliminating the source of vertical instability by compensating $B_R$ field at the center of plasma currents in each steps as following.
\begin{equation}
I_{\text{feedback}}=-g_z \genfrac{.}{|}{}{} {B_{R,\text{vacuum}}} {B^*_{R,\text{feedback}}}
  _{ R_{\text{cur}},Z_{\text{cur}} }
\label{eq:general_zp}
\end{equation}
where the minus sign indicates a compensation, $g_z$ is an adjustable constant, and $B_{R,\text{vacuum}}$ and $B^*_{R,\text{feedback}}$ are the radial magnetic fields by external equilibrium conductor currents (i.e. vacuum field) and by unit currents of vertical stabilizing coils, respectively.
Also note that there is no $I_P$ dependency in this method.
If $g_z=1.0$, the exactly same $B_R$ field is compensated by the feedback currents.
In TES, practically $2.0 \le g_z \le 2.5$ is used to ensure a general stabilization by using a up-down symmetric pair of coils, which is set to be located radially in the middle of and vertically just outside computational domain.
For the evaluation of $B_R$, we use the effective current center ($R_{\text{curr}}, Z_{\text{curr}}$) instead of the magnetic axis for a better description of axisymmetric plasma motion as following
\begin{eqnarray}
R^2_{\text{cur}} &=& \dfrac{1}{I_p} \mathop{ \int } R^2 J_{\phi,pl}(R,Z)d\Omega_{pl} \\
Z_{\text{cur}} &=& \dfrac{1}{I_p} \mathop{ \int } Z J_{\phi,pl}(R,Z)d\Omega_{pl} \nonumber
\label{eq:curr_center}
\end{eqnarray}
Note that a generalized method for the radial stabilization is not described here (due to lack of practical interest) but also possible in a similar way.

\subsection{Validation of vertical stability and its stabilization}

\begin{figure}[!hbtp]
\centering
\includegraphics[width=0.48\textwidth]{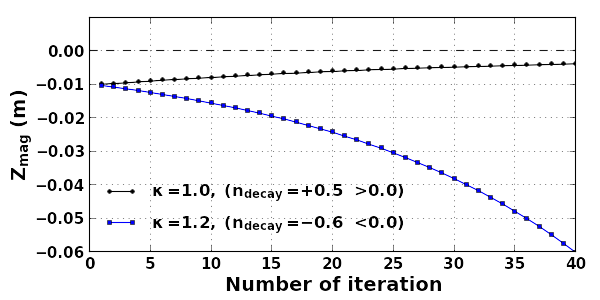}
\includegraphics[width=0.48\textwidth]{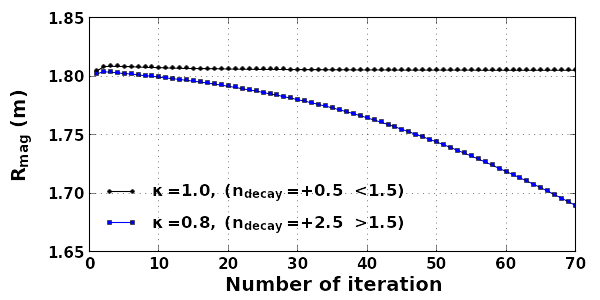}
\caption{(Color online) Comparisons of plasma displacement responses to the initial perturbations of $\Delta Z=-0.01$ m and $\Delta R=-0.01$ m for the stability test.
Two equilibria with $\kappa=1.0$ and $\kappa=1.2$ are tested for the vertical stability (left) and another two equilibria with $\kappa=1.0$ and $\kappa=0.8$ are tested for the radial stability (right). 
}\label{fig:stability_check}
\end{figure}
For a validation of the generalized stabilization method described above, we first test the validity of force-balance relation solved in TES, by considering the natural axisymmetric instability.
From FIG. \ref{fig:ndecay}, it is obvious that the equilibrium with $\kappa=1.2$ is expected to be vertically unstable ($n_{\text{decay}}<0$), while the equilibrium with $\kappa=1.0$ to be stable ($n_{\text{decay}}>0$), if the force-balance relation in TES is correct.
Similarly, the equilibrium with $\kappa=0.8$ is expected to be radially unstable ($n_{\text{decay}}>1.5$), while the equilibrium with $\kappa=1.0$ to be stable ($n_{\text{decay}}<1.5$).
Remind that these equilibria were obtained by specifying the plasma boundary and thus produced the required external equilibrium fields as a result (i.e. as a {\it semi free} boundary problem).
To test the natural vertical stability without any additional stabilization, the equilibrium analysis is re-performed as an {\it ideally free} boundary problem, i.e. by specifying the external equilibrium coil currents which were obtained from the FIG. \ref{fig:ndecay}.
As a seed for vertical or radial instability, a small perturbation is added into the initial position of plasma boundary, which is used in 0-th Picard iteration.

The comparisons of vertical and radial displacement responses to small deviations of $\Delta Z=-0.01$ m and $\Delta R=-0.01$ m are shown in FIG. \ref{fig:stability_check}.
On the left, the vertical stability is tested by an initial perturbation, $\Delta Z=-0.01$ m, for two equilibria; one with $\kappa=1.0$ (black dotted line) and the other with $\kappa=1.2$ (blue dotted line).
Consistently with the theoretical expectations, the initial perturbation of the former was naturally stabilized, while the one of the latter was exponentially diverged.
On the right, the radial stability is tested by an initial perturbation, $\Delta R=-0.01$ m, for two equilibria; one with $\kappa=1.0$ (black dotted line) and the other with $\kappa=0.8$ (blue dotted line).
Similarly, the initial perturbation of the former was naturally stabilized or stable, while the one of the latter was exponentially diverged.
Therefore, it confirms that the force-balance relation used in TES is correctly solved and the associated instability is precisely consistent with the theory.

\begin{figure}[!htbp]
\centering
\includegraphics[width=0.3\textwidth]{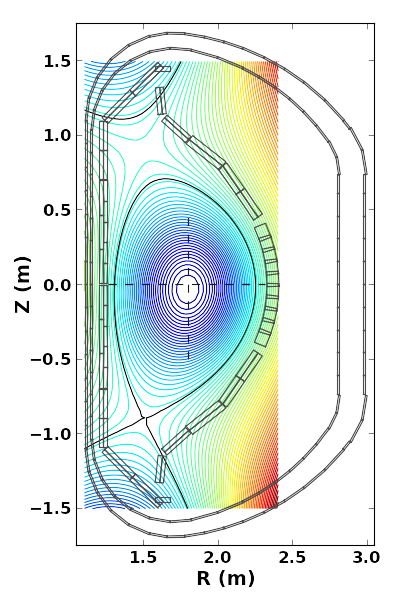}
\includegraphics[width=0.6\textwidth]{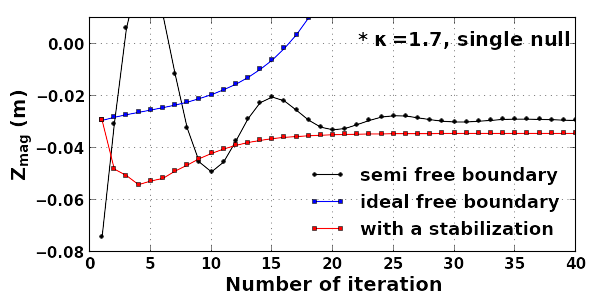}
\caption{(Color online) Vertical displacement responses to an initial perturbation of $\Delta Z=-0.03$ m are compared for the equilibrium shown on the left.
The first one (black) is during TES iteration as a {\it semi free} boundary problem, the second one (blue) as an {\it ideally free} boundary problem, and the third one (red) with the generalized vertical stabilization.
}\label{fig:general_zp}
\end{figure}

For the validation of the generalized stabilization method, a further strongly shaped and up-down asymmetric plasma is considered as a worst case.
The reference equilibrium is obtained in a single null (SN) configuration with $I_P=-0.5$ MA, $B_T=2.5$ T, $\beta_P=0.1$, and $\kappa=1.7$ ($n_{\text{decay}}=-1.8$) as shown in FIG. \ref{fig:general_zp}.
The comparison of vertical displacement responses with and without the generalized stabilization is shown on the right of the FIG. \ref{fig:general_zp}.
The evolution of a {\it semi free} boundary solution (black line) shows that it converged to $Z_{\text{mag}}=-0.03$ m (thus it is a reference equilibrium position).
In case of {\it ideally free} boundary solution (blue line) without any stabilization, it was slowly drifted upward and finally diverged, as expected.
Then, by applying the generalized vertical stabilization (red line), the evolution was really stabilized so that it was smoothly evolved and converged to the reference position closely.
Here, the final difference of vertical position compared with the reference is about 0.5 cm.
Therefore, it demonstrates that the generalized method can effectively stabilize the natural vertical instability of elongated plasmas and automatically guide the plasma to an equilibrium position, that is consistent with that from a {\it semi free} boundary solution.

\section{Extension to Advanced Equilibrium Analysis}
\label{sec:extension}

Recently new types of tokamak equilibria have been proposed and studied in various devices, in order to resolve the issue of an excessive heat and particle fluxes onto the plasma facing components in ITER and beyond. 
These are featured by a new divertor configuration such as snowflake \cite{Ryutov2007PoP} and (super) X divertors \cite{Kotschen2007PoP}.
Particularly the snowflake equilibrium requires to have a second-order zero of poloidal flux at the null-field point so that it is not straight-forward to deal with it by a conventional free boundary equilibrium solver \cite{Lackner2013FST}.
To solve this new equilibrium with specified plasma boundary, the minimization constraint, Eq. (\ref{eq:eq_pf}), for required external equilibrium field currents is modified in TES as follows

\begin{eqnarray}
\label{eq:eq_pf_sf}
\min_{\Delta I_{\text{coil}}}
\left[ 
\sum^{N_{\text{bndry}}}_{j=1} { \left\lbrace \sum^{N_{\text{coil}}}_{i=1} 
  \Big( G(R_j,Z_j;R_i,Z_i)\cdot \Delta I_{\text{coil},i} \Big)
 -\Delta \psi(R_j,Z_j) \right\rbrace^2 } \right. & \nonumber \\
+\sum^{N^{\text{SF}}_{\text{Xpt}}}_{j=1} { \left\lbrace   
  \sum^{N_{\text{coil}}}_{i=1} 
  \Big( \genfrac{}{}{}{} {\partial {G_{B_R}(R_j,Z_j;R_i,Z_i)}}{\partial Z} \cdot \Delta I_{\text{coil},i} \Big)
 -\genfrac{.}{|}{}{} {\partial B_{R}}{\partial Z} _{(R_j,Z_j)}
 \right\rbrace^2 } & \nonumber \\
+\sum^{N^{\text{SF}}_{\text{Xpt}}}_{j=1} { \left\lbrace 
  \sum^{N_{\text{coil}}}_{i=1} 
  \Big( \genfrac{}{}{}{} {\partial {G_{B_Z}(R_j,Z_j;R_i,Z_i)}}{\partial R} \cdot \Delta I_{\text{coil},i} \Big)
 -\genfrac{.}{|}{}{} {\partial B_{Z}}{\partial R} _{(R_j,Z_j)}
  \right\rbrace^2 } & \nonumber \\
\left.
+\gamma^2 \sum^{N_{\text{coil}}}_{i=1}{ \Big( \Delta I_{\text{coil},i} \Big)^2 }
\right]
\end{eqnarray}
where $\genfrac{}{}{}{0} {\partial B_{R}}{\partial Z} = \genfrac{}{}{}{0} {\partial}{\partial Z} \left( -\genfrac{}{}{}{0} {1}{R}\genfrac{}{}{}{0} {\partial \psi}{\partial Z} \right)=-\genfrac{}{}{}{0} {1}{R} \genfrac{}{}{}{0} {\partial^2 \psi}{\partial Z^2}$ and similarly $\genfrac{}{}{}{0} {\partial B_Z}{\partial R}=+\genfrac{}{}{}{0} {1}{R} \genfrac{}{}{}{0} {\partial ^2 \psi}{\partial R^2}$.
By using this, the snowflake equilibrium that requires a second-order zero of $\psi(R,Z)$ can be directly obtained without any special treatment in TES.

\begin{figure}[!htbp]
\centering
\includegraphics[width=0.6\textwidth]{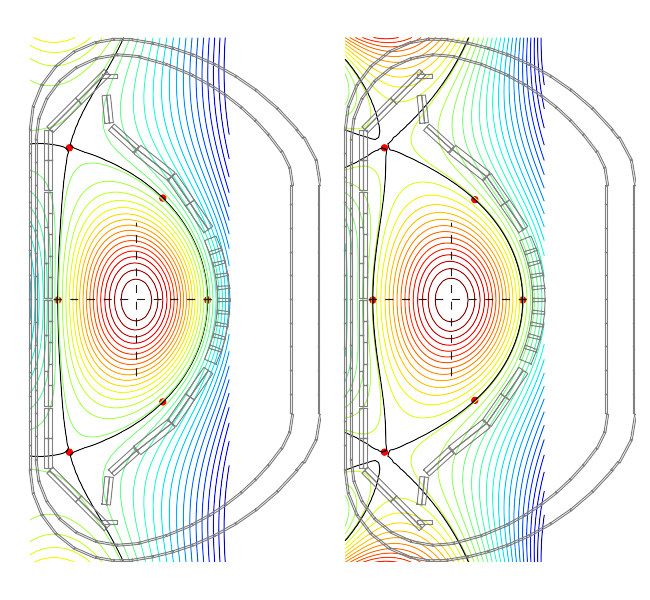}
\caption{(Color online) A comparison of two distinctive plasma equilibria. One is with a typical double null divertor (left) and the other with a snowflake divertor (right). The plasma boundary points specified in calculation are marked with a red circle.
}
\label{fig:DN_SF}
\end{figure}
Figure \ref{fig:DN_SF} shows a comparison of two equilibria obtained by TES with identical plasma equilibrium parameters, which are $I_P=$1.0 MA, $B_T=$2.5 T, $\kappa=$2.0, and $\beta_P=$0.2.
One (on the left) is a typical double null (DN) divertor and the other (on the right) is a snowflake (SF) divertor configurations.
The difference of two equilibria is easily seen from the magnetic distributions around the field null points.
In the SF configuration, it is clearly seen that three concave and another three convex distributions are formed alternately, centred at the up-down symmetric field null points, i.e. a second order zero of poloidal magnetic flux is formed.

\begin{table}
\caption{Required external coil currents (kA) to form the equilibria shown in FIG. \ref{fig:DN_SF}}
\begin{ruledtabular}
\begin{tabular}{ccc}
 & Double-Null (DN) & Snow-Flake (SF) \\
\colrule
PF1  & -6.02  &   -1.04  \\
PF2  &  9.64  &  -19.05  \\
PF3  &  6.10  &  135.27  \\
PF4  &  7.07  &  -67.32  \\
PF5  &  8.33  &   18.60  \\
PF6  & -1.03  &   -5.30  \\
PF7  & -7.26  &   -4.03  \\
\end{tabular}
\end{ruledtabular}
\label{tab:Ieq_SF}
\end{table}
Table \ref{tab:Ieq_SF} shows the external equilibrium coil currents required to form the target equilibria shown in FIG. \ref{fig:DN_SF}.
It is important to note that in the case of SF equilibrium some of coil currents are required extremely large values, while in the case of DN equilibrium all coil currents are well balanced.
It indicates that it is not practically possible to form the SF equilibria onto the KSTAR by using current coil system.
Therefore, a new coil system, specially designed for SF divertor, is essentially needed.
In fact, it is consistent with the recent highlighted issue \cite{Lackner2013FST} in the study on advanced divertor configurations.
In addition, it is worthwhile to note that the SF equilibrium here is solved self-consistently by considering full force-balance relations in a toroidal system, while in the reference {\cite{Lackner2013FST}, it is solved by using a simplified wire plasma model.

\section{CONCLUSIONS}
\label{sec:conclusion}
A free-boundary tokamak equilibrium solver, developed for advanced study of tokamak equilibra, was described with various validation results.
The developed solver, named as TES, is characterized by two distinctive features.
At first, a generalized stabilization method for intrinsic axisymmetric instabilities was applied, which is encountered after all in equilibrium calculation under a free boundary condition.
In this method, the source of axisymmetric instabilities is directly removed or minimized, instead of feedback controlling the plasma position to a target location.
Thus, it ensures in general that the TES code produces a solution stably even under highly (axisymmetrically) unstable conditions.

The other important feature is an extension to deal with a new divertor geometry such as snowflake or X divertors. 
To deal with the innovative divertor concept, particularly the snowflake divertor, the equilibrium solver needs to be able to control the location of second order zero of poloidal magnetic field.
By implementing this functionality into the TES code, it was demonstrated that the snowflake type of advanced tokamak equilibria can be analysed in consideration of full toroidal force balance relations, instead of using a simplified wire plasma model.

For the validation of TES code, the uniqueness of a solution was confirmed by the independence on variations of computational domain, the mathematical correctness and accuracy of equilibrium profiles were checked by a direct comparison with the generalized Solov'ev equilibrium, and the governing force balance relation was tested by examining the intrinsice axisymmetric instabilities.

As a valuable application, a snowflake equilibrium was analysed by taking into account the KSTAR equilibrium coil system. 
Since the KSTAR has a limited set of equilibrium control coils, it is important to check whether the innovative divertor equilibria can be realized in the current system.
The analysis results suggest that practically it is not possible to form a snowflake equilibrium in current KSTAR device so that additional control coils need to be considered for the study of advanced divertors in future.


\begin{acknowledgments}

This work was supported by the Korean Ministry of Science, ICT and Future Planning under the KSTAR project contract.
\end{acknowledgments}

\end{document}